# Quantifying magma mixing with the Shannon entropy: application to simulations and experiments

Perugini D.[1]*, De Campos C.P.[2], Petrelli M.[1], Morgavi D.[1], Vetere F.P.[1] & Dingwell D.B.[2]

[1] *Department of Physics and Geology, University of Perugia, Piazza Università, Perugia 06100, Italy*

[2] *Department of Earth and Environmental Sciences, Ludwig-Maximilian-University, Theresienstrasse 41, 80333, Munich, Germany*

*\* Corresponding author*:

Diego Perugini (e-mail: diego.perugini@unipg.it)

# Abstract

We introduce a new quantity to petrology, the Shannon entropy, as a tool for quantifying mixing as well as the rate of production of hybrid compositions in the mixing system. The Shannon entropy approach is applied to time series numerical simulations and high-temperature experiments performed with natural melts. We note that in both cases the Shannon entropy increases linearly during the initial stages of mixing and than saturates towards constant values. Furthermore, chemical elements with different mobilities display different rates of increase of the Shannon entropy. This indicates that the hybrid composition for the different elements is attained at different times generating a wide range of spatio-compositional domains which further increase the apparent complexity of the mixing process.

Results from the application of the Shannon entropy analysis are compared with the concept of Relaxation of Concentration Variance (RCV), a measure recently introduce in petrology to quantify chemical exchanges during magma mixing. We derive a linear expression relating the change of concentration variance during mixing and the Shannon entropy.

We show that the combined use of Shannon entropy and RCV provides the most complete information about the space and time complexity of magma mixing. As a consequence, detailed information about this fundamental petrogenetic and volcanic process can be gathered. In particular, the Shannon entropy can be used as complement to the RCV method to quantify the mobility of chemical elements in magma mixing systems, to obtain information about rate of production of compositional heterogeneities, and to derive empirical relationships linking the rate of chemical exchanges between interacting magmas and mixing time.

# 1. Introduction

The mixing of magmas has been a subject of considerable scientific research in the last three decades (Anderson, 1982; Russell, 1990; Bateman, 1995; Abe, 1997; Jellinek and Kerr, 1999; Bergantz, 2000; Perugini et al., 2003; Slaby et al., 2011; Perugini and Poli, 2012; Pietruszka et al., 2015; Shorttle, 2015). It is considered by many as a major process generating extreme compositional variations in rock suites as well as one of the main processes responsible for triggering highly explosive volcanic eruptions (Sparks et al., 1977; Murphy et al., 1998; Leonard et al., 2002; Martin et al., 2008; Perugini et al., 2010; Tomiya et al., 2013). Magma mixing has been investigated using several different approaches: from classical geochemical studies (Langmuir et al., 1978; Poli et al., 1996; Xu et al., 2014; Hagen-Peter et al., 2015), through numerical simulations (Oldenburg et al., 1989; Folch and Martí, 1998; Jellinek and Kerr, 1999; Petrelli et al., 2011) and experiments with both synthetic and natural compositions (Kouchi and Sunagawa, 1985; Morgavi et al., 2013a; Perugini et al., 2013; Laumonier et al., 2014). These studies highlight that the mixing of magmas is characterized by an extreme compositional variability in both space and time, resulting from the development of chaotic mixing processes between the interacting melts (Flinders and Clemens, 1996; Perugini et al., 2003; De Campos et al., 2011; Slaby et al., 2011). Deviations from classic linear mixing trends have been documented in natural samples, numerical simulations and high-temperature experiments (Perugini et al., 2006; De Campos et al., 2011; Morgavi et al., 2013a). These observations underline the non-linearity of this natural process. And this fundamental nonlinearity makes magma mingling/mixing one of the most complex petrogenetic processes on our planet.

Despite the considerable amount of literature on the subject, many fundamental aspects of magma mixing still remain unresolved. In particular, the time evolution of magma mixing still requires study in order to understand the impact of mixing upon the generation of extremely variable compositional domains in space and the rate of production of hybrid compositions with respect to mixing dynamics. This latter issue is of paramount importance because the rate at which hybrid compositions are generated is directly related to the disappearance of end-member compositions. The ability of a magmatic system to preserve information about the end-member compositions is crucial in both petrology and volcanology. For example, identifying the original compositions of the end-members involved in the mixing processes is essential for reconstructing the geochemical characteristics of source regions, with profound implications for geodynamics inferences based upon magma chemistry.

Here, in order to contribute to a better understanding of the time evolution of magma mixing processes and the rate at which hybrid compositions are generated, we introduce a quantity new to geosciences, the Shannon entropy, as an additional petrological tool to quantify the rate of production of hybrid compositions in a mixing system. The Shannon entropy analysis is applied to previously published numerical simulations and high-temperature experiments performed with natural melts (Morgavi et al., 2013a; Perugini et al., 2013). We also compare this quantity with the concept of Relaxation of Concentration Variance (RCV), recently introduced in petrology to quantify magma mixing (Morgavi et al., 2013a; Perugini et al., 2013). We show that Shannon entropy can be used a complementary petrological tool that, together with RCV, can provide a comprehensive view of the complexity in space and time of magma mixing processes.

## 2. Numerical simulations and high-temperature experiments with natural melts

Here we briefly recall some basic information about the numerical simulations and the high-temperature experiments that we subject in this work as tests of the applicability of the Shannon entropy analysis to magma mixing systems. Those interested in the details of the numerical methods or the high-temperature experiments can find full methodological descriptions in Morgavi et al. (2013a) and Perugini et al. (2013).

*2.1. Numerical simulations*

The numerical simulations are based on a numerical scheme that has been used to reproduce, as a first approximation, mixing structures and compositional patterns analogous to those generated during the mixing of natural melts in the volcanic environment (Wada, 1995; De Rosa et al., 2002; Perugini et al., 2003). The system describes the basic elements characterizing chaotic mixing systems, including magma mixing: i) advection, i.e. stretching and folding of fluid elements and ii) mass exchange via chemical diffusion (Flinders and Clemens, 1996; Perugini et al., 2003; De Campos et al., 2011). This advection/diffusion numerical system was proposed as a proxy to simulate magma mixing processes between degassed and crystal-free melts (Perugini et al., 2003; Morgavi et al., 2013a).

The stretching and folding dynamics were triggered by applying the "sine-flow" chaotic mixing system, which is a well know prototypical chaotic flow (Liu et al., 1994; Clifford et al., 1999, 1998). As shown by (Perugini et al., 2003), in order to make the system more similar to a natural scenario, the flow is continuously re-modulated by changing randomly the control parameter of the flow (i.e. parameter $k$ in the works of ( Perugini et al., 2004; Morgavi et al., 2013a). Chemical diffusion was simulated using a finite difference scheme as reported in (Perugini et al., 2003).

In the simulations two magmas were considered, A (light color) and B (dark color) (see below). The B magma is constituted by eight elements having the same initial concentration equal to 250 (in grey levels), but different diffusion coefficients (*D*). *D* varies from 0.0 to 1.0. The initial concentration of elements in the A magma is set to 20 (in grey levels). This configuration is a choice of convenience in the calculations and is adopted in order to introduce chemical gradients of the elements between the two magmas.

The chaotic mixing simulations were carried out starting from the initial configuration reported in Fig. 1a. As an example, Fig. 1b-d show snapshots of the mixing system at different times for an element with a diffusion coefficient *D*=0.3. The figures show that, as the mixing time increases, the system becomes progressively more "blurred". This is due to the fact that the time evolution of stretching and folding dynamics generates a continuous increase of contact interfaces between the two liquids which inevitably results in an enhancement of chemical exchanges.

Simulations were performed for 100 iterations of the numerical scheme. The variation in time of concentration of elements across a transect ($\alpha$–$\beta$) passing through the mixing system (Fig. 1) is shown in Fig. 2 for an element with *D*=0.3. The graphs show that, as mixing time progresses, the system becomes progressively more homogeneous and the chemical variability is drastically reduced. At large mixing times the entire compositional spectrum along the transect eventually collapses to a single composition corresponding to the hybrid of the system (Fig. 2d).

Fig. 3 shows the tenth iteration of the mixing system for three elements having diffusion coefficients (*D*) values equal to 0.05, 0.3 and 1.0. The pictures on left-hand panels show that, at the same time, as the value of *D* increases, the mixing system simultaneously becomes progressively more "blurred" due to the increasing diffusive fluxes of the different elements. The graphs of the right-hand panels report the compositional variations of the elements along the transect. It is evident that the different elements define very different compositional patterns. In particular, the element with the lowest diffusivity (*D*) shows a variation including the two end-member compositions with a few data points displaying a compositional shift towards the hybrid composition. As the *D* value increases, the compositional pattern becomes progressively less variable with an increasing number of data points falling in the field of the hybrid composition (Fig. 3b-c). Clearly then, in the same mixing system and at the same time, different chemical elements can experience very different histories, with some elements still showing large variability whereas others are almost completely homogenized.

*2.2. High-temperature experiments with natural melts*

Experiments were performed in a high temperature centrifuge furnace (Dorfman et al., 1996; Veksler et al., 2007) using an alkali-basaltic and a phonolitic end-member from the Campi Flegrei (Italy) volcanic area (de Vita et al., 1999; Di Vito et al., 1999; Arienzo et al., 2009). Below, the phonolitic and the alkali-basaltic melt are referred to as felsic and mafic melt, respectively. Full details of the experimental conditions are reported in (Perugini et al., 2013) and some essential aspects are repeated here.

The end-members glasses were mounted into a sealed Pt-capsule, in a buoyancy unstable geometry (phonolite in the outer part "bottom" and alkali-basalt in the inner part "top"), and then introduced into the outer container of the centrifuge. The same procedure and geometry were used in all runs. All experiments were performed at 1,200°C and an acceleration of $10^3$ g (g = 9.81 m/s$^2$). The only independent variable was run duration with three experiments at 5, 20 and 120 minutes being performed. During quenching, temperature inside the sample cooled from 1200 to 1100 °C in 10 s, from 1100 to 1000 °C in 15 s, and further cooling of the sample to 800 °C took about 1.5–2.0 min. This cooling yielded completely glassy samples, as confirmed by optical and scanning electron microscope examination.

By forcing the mafic melt towards the outside of the capsule ("down"), the felsic melt is driven "up" to the other side. During the motion of the mafic melt, parts of it were entrained into the felsic melt triggering a mixing process. During rotation of the capsule, those parts of mafic melt entrained into the felsic one were pulled towards the mafic magma by centrifugal force and mixing processes was largely localized at the interface between the two melts.

Experimental samples from 5, 20 and 120-minute experiments are shown in Fig. 4. The sample quenched after 5 min exhibits clear evidence of the mafic, denser melt, moving to the side opposite its initial location ("down") due to the centrifugal force. During this motion, tendrils of mafic melt were entrained in the felsic one producing filament-like patterns extending to the top of the experimental sample (Fig. 4a and d). The experimental sample at 20 min displays a clear evolution of the system towards a higher mixing efficiency (Figs. 4b and e). In particular, the filaments occurring in the upper part of this sample collapsed towards the interface between the two melts. In this region vortex-like structures are generated by vigorous stretching and folding dynamics producing an intricate pattern of alternating filaments of the two melts (Figs. 4b and e). After 120 min the experimental sample appears largely divided into two zones (Figs. 4c and f) populated by the two melts: mafic in the lower part and felsic in the upper part. A few shades of filaments of the mafic melt are still visible in the middle-upper part of the sample. From optical analysis, hence, the time evolution of the mixing system indicates an increasing mixing progress passing from 5 to 120 min with a strong smoothing of color gradients due to the increasing cumulative diffusive interaction between the starting end-members, analogous to the results of numerical simulations.

As most of the mixing process occurred along the contact interface between the two melts, compositional transects consisting of ca. 1000 data points were analyzed across these interfaces (paths shown along the dotted lines; Fig. 4d-f). Concentrations of major elements were determined by electron microprobe (Perugini et al., 2013).

The variation of some representative major elements at different mixing times along the three transects is shown in Fig. 5. In the figure, element concentrations of both the end-members and the hybrid melt are also marked, in order to follow better the evolution of the mixing process (Perugini et al., 2013). The time evolution of compositional variability along the transect indicates a progressive increase of mixing intensity due to the combined action of stretching and folding dynamics, the consequent increase of contact interfaces between the two melts, and the onset of chemical diffusion (Fig. 5a-c). This is demonstrated by the fact that the fluctuations of concentration of chemical elements decrease continuously passing from the 5 min to the 120 min experiment (Fig. 5). As the mixing time increases the compositional variability is progressively reduced and shifted towards the hybrid composition.

## 3. Quantitative analyses of mixing using the Shannon entropy

Generically, mixing refers to a process that reduces compositional heterogeneity. Entropy is the rigorous measure of disorder or system homogeneity. Thus, in this work we will explore how to employ the concept of Shannon entropy (Shannon, 1948) to characterize the state of mixing in the previously documented numerical simulations and high-temperature experiments. In particular, the concept of Shannon entropy is applied here to measure the intensity of mixing and its variation in time. As we are interested in the mixing intensity, we have redrawn the compositional time series arising from the numerical simulations as well as from the experiments, focusing on the number of data points that attained the hybrid composition during mixing. This approach is based on the idea that the higher the number of hybrid analyses in the system, the larger the mixing intensity.

Specifically, in order to apply the Shannon entropy analysis, we redraw the compositional series arising from the numerical simulations (Fig. 2-3) and high-temperature mixing experiments (Fig. 5), as point distributions along a line (1D point distribution).

For each compositional series, the number of data points falling within the field of the hybrid composition is monitored. When the concentration of a given element attains the hybrid composition, a point is drawn on a line at the relative position of the data point in the compositional series. To better visualize this process, we represent the point-distribution of the hybrid composition as a compositional "barcode" (BC). Fig. 6 illustrates the method graphically. For example, Fig. 6a exhibits a compositional time series generated by the numerical simulation. The location of the hybrid composition is marked as a gray area to include a 10% compositional variability. Data points having attained the hybrid composition (i.e. those falling in the grey area) are white. Fig. 6b shows the point distribution of the hybrid compositions along a 1D line. For purpose of visualization the representation is rendered as a bar-code. In particular, as noted above, a segment of arbitrary length is drawn at the location of each hybrid data point resulting in a bar-code representation of the hybrid data points distribution along the compositional transect (Fig. 6c). Note, however, that the Shannon entropy analysis which follows is nevertheless performed considering the point distributions themselves.

The point distributions (compositional "bar-codes", BC) extracted from the mixing systems (numerical simulations and experiments) were used to calculate the evolution of the Shannon entropy ($S$) (Shannon, 1948; Camesasca et al., 2006) of the mixing system in time. In particular, we consider the transect to be divided up into a grid made up of a number $M$ of cells $c_i$ with area $a_i$, such that $\sum_{i=1}^{M} a_i = A$, the total area of the transect (Fig. 7). As shown by (Baranger et al., 2002), $S$ can then be calculated using the following equation

$$S(t) = -\sum_{i=1}^{M} p_i(t) \log p_i(t) \qquad [\text{Eq. 1}]$$

where $p_i(t)$ is the probability that hybrid data points fall into cell $c_i$ at time $t$. In practice $p_i(t)$ is calculated for each cell as the ratio between the number of points present in the cell ($N_c$) and the total number of points in the compositional transect ($N_{tot}$). Thus, $S(t)$ is calculated by summing the values of [$p_i(t) \log p_i(t)$] calculated, considering all cells $M$. The "worst" mixing scenario, corresponding to the state of the system before starting the mixing process, is characterized by $S=0$ (i.e.: no data point attained the hybrid composition; Fig. 7) and the best mixing (all cells contain equal number of data point with the hybrid composition; Fig. 7) is characterized by the maximum entropy $S_{max}$. In order to compare the behaviour of the Shannon entropy ($S$) for different elements, $S$ values have been normalized ($S_N$) to the maximum value ($S_{max}$) that the entropy would attain when the system is completely populated by hybrid compositions. This way $S_N$ values are bounded between 0.0 and 1.0. $S$ is a quantity that increases with time, and it is related to the process governing the evolution of the system (Baranger et al., 2002).

The cell size chosen to quantify mixing intensity via the Shannon entropy ($S$) defines the scale of observation (the "magnifying glass") at which one looks at the system. The larger the cell size is, the larger the scale of observation at which the observer evaluates the Shannon entropy ($S$). Since there is no established method to determine the size of the cells to be used for the estimation of $S$, we attempted to calculate $S$ on the same compositional transect using several cell sizes. Results show that, with increasing cell size, the value of $S$ decreases. This agrees with the fact that larger cells correspond to lower values of $p_i$ and, hence, to lower values of $S$. Thus, $S$ is not an absolute value, but depends on the cell size used to calculate it. However, it is worth noting that the decrease in $S$ with increasing cell size is systematic as $S$ values are shifted proportionally maintaining their relative values. As we are interested here in the relative variation of $S$ with the time progression of mixing, we used a cell size of 5 pixels and 50 μm for the compositional time series of the simulations and experiments, respectively. Recall that on the basis of the above discussion, changing cell size will result in a proportional scaling of all $S$ values and the overall structure of the results will be maintained.

*3.1. Application of Shannon entropy to numerical simulations*

As for the numerical simulation, given the initial proportions of magma A and B (60% magma and 40% of magma B), the hybrid composition corresponds to grey level 158. Rather than fixing threshold at this value, we decided to introduce a tolerance of 10% (i.e. gray scale values 142-174) in defining the hybrid composition. This choice was considered appropriate taking into account the analytical errors typically associated with chemical analyses.

The bar-code representation of the hybrid data point distribution for the transects extracted from the numerical simulations are reported on the top of the compositional series shown in Fig. 2 and 3. These plots show that, as mixing time progresses, an increasingly larger number of hybrid compositions are generated, resulting in a progressively more populated bar-code plots (Fig. 2). Further, at the same time, chemical elements with different mobilities show different bar-code plots. In particular, the higher the mobility of the element is, the larger the number of data points having attained the hybrid composition (Fig. 3).

The graph in Fig. 8 reports the variation of the normalized Shannon entropy ($S_N$) against mixing time (i.e., number of iterations of the simulated mixing system, $t$). From the graph it can be seen that $S_N$ increases with increasing mixing time for all simulated elements. At longer mixing times $S_N$ tends toward a time-invariant value of one corresponding to the completely homogenized mixing system (i.e. the mixing system consists entirely of the hybrid composition). This behaviour is observed for all elements, with the distinction that, as $D$ increases, the slope ($dS/dt$) of the rectilinear segments of the curves also increases (Fig. 8). In order to quantify the different increase of entropy with time for the different chemical elements, the linear portions of the curves in Fig. 8 were fitted by linear regression and the slope coefficient (i.e.

the rate of increase of the Shannon entropy, $dS/dt$) for each simulated element was determined. Results show an increase of $dS/dt$ from ca. 0.021 to 0.078 as the diffusion coefficient of the element increases.

*3.2. Application of Shannon entropy to high-temperature experiments*

The Shannon entropy for the experiments was calculated from compositional series measured along the transects (Fig. 5). As was performed for the case of numerical simulations, in order to facilitate comparisons among the different chemical elements, $S$ values were normalized ($S_N$) by the maximum value ($S_{max}$) that the entropy would attain when the system is completely populated by hybrid compositions (i.e. all concentrations along the transect attain the hybrid composition). In the case of magma mixing experiments the value of the hybrid concentration for each chemical element was determined using the method proposed by (Perugini et al., 2004; 2013) and based on the study of compositional histograms. Values of the hybrid concentration for the major elements considered here are taken from (Perugini et al., 2013).

The graph in Fig. 9 reports the variation of the normalized Shannon entropy ($S_N$) against magma mixing time ($t$). As for the case of numerical simulations, $S_N$ increases with increasing mixing time for all simulated elements. At longer mixing times (i.e. 120 min.) $S_N$ tends to saturate toward constant steady-state values. Although this behaviour is observed for all measured major elements, the different elements show different slopes ($dS/dt$) of the rectilinear segments of the curves (i.e. excluding the data points at 120 min; Fig. 9). In order to quantify the different increase of entropy with time for the different chemical elements, the linear portions of the curves in Fig. 8 were fitted by linear regression, as for the numerical simulations, and the slope coefficient (i.e. the rate of increase of the Shannon entropy, *dS/dt*) for each chemical element was determined.

**4. Shannon entropy vs. Relaxation of Concentration Variance (RCV)**

Recently, Morgavi et al. (2013a) and Perugini et al. (2013) we have introduced the concept of Relaxation of Concentration Variance (RCV) as a tool to quantify the mixing intensity. This concept relies upon the concept of concentration variance, a quantity commonly used in the fluid dynamics literature (Rothstein et al., 1999; Liu and Haller, 2004) to evaluate the degree of homogenization of fluid mixtures. In general, concentration variance $\sigma_n^2$ quantifies the spread of compositional variation in a mixing system. In other words, at increasing mixing time the concentration variance decreases exponentially towards zero meaning that the system becomes progressively more homogeneous trending towards the hybrid composition. In chaotic mixing systems, such as those considered in this work (i.e. the numerical simulations and the experiments), the concentration variance decreases exponentially with time (Morgavi et al., 2013a; Perugini et al., 2013). As an example, Fig. 10 shows the variation of concentration variance with time for some representative chemical elements for both the numerical simulations and the high-temperature experiments. The plots show that $\sigma_n^2$ decays quickly during the early stages of mixing and then shows a relaxation towards $\sigma_n^2 = 0$ for long mixing times. The rate of decay of concentration variance is different for the different elements. This is quantified by fitting the data on the plots of Fig. 10 by an exponential equation such as

$$\sigma_n^2(C_i) = C_0 \cdot \exp(-Rt) + C_1 \qquad [\text{Eq. 2}]$$

where $C_0$, $R$ and $C_1$ are fitting parameters and $t$ is the mixing time. From Eq. [2] it is clear that the rate of concentration decay can be expressed in terms of the parameter $R$. This parameter was termed by Perugini et al. (2013) "Relaxation of Concentration Variance" (RCV) and is a measure of the mobility of elements in the mixing system.

The plot of Fig. 11 displays the variation of Shannon entropy ($dS/dt$) versus RCV for the numerical simulations. A linear positive correlation between the rate of increase of Shannon entropy with time ($dS/dt$) and RCV is apparent. In particular, the values of $dS/dt$ increase as the mobility of chemical elements, quantified by the RCV, increases.

Fig. 12 shows the variation of Shannon entropy ($dS/dt$) against RCV for the major elements measured on the high-temperature experiments. As for the case of numerical simulations the plot shows a linear positive correlation between these two variables. In particular, the highest values of $dS/dt$ are recorded for Na followed by Al, Mg, K, Ca, Ti, Si and Fe. This sequence matches perfectly the sequence of element mobility defined on the basis of the RCV method (Perugini et al., 2013).

From the results presented above it is clear that the rate of increase of Shannon entropy is different for different chemical elements, each having different mobility in the mixing system. In particular, our results indicate that elements having larger mobility display higher rate of production of hybrid compositions. This is the origin of the ability of Shannon entropy to identify the velocity with which the different chemical elements converge towards the production of hybrid compositions in the mixing system. Here, we have shown that a linear relationship exists between the Shannon entropy and the RCV. These results greatly enhance our ability to deal with the complexity of magma mixing processes and, in particular, our ability to reconstruct the original compositions of interacting melts and to identify the potential hybrid composition that the system would eventually attain. These aspects are discussed in the following section.

**4. Discussion and conclusions**

The new measure introduced for magma mixing, the Shannon entropy (S), allows us to study in detail the interplay between magma mixing dynamics and the generation of hybrid compositions. The Shannon entropy is able to take into account, in a single variable, all possible factors that can play a role in the evolution of the magma mixing system towards the final hybrid composition. We have demonstrated that S represents a new tool to estimate the velocity at which hybrid compositions are generated for each chemical element.

The results presented here support and extend previous studies (Morgavi et al., 2013a; Perugini et al., 2013) where it was shown, by means of the Relaxation of Concentration Variance (RCV) approach, that in the same magma mixing system, at the same time, different degrees of homogenization are possible for the different chemical elements. Now, the use of Shannon entropy allows us to shed new light about the rate of production of hybrid compositions in magmatic systems and offers the opportunity to penetrate even more deeply into the mechanisms acting during interaction of silicate melts.

In particular, the analysis of the Shannon entropy can be complementary to the RCV method when end-member compositions are difficult to be recognized. In fact, while the conceptual model of RCV is a proxy for the relative mobility of the different chemical elements, it also requires the knowledge of the starting compositions of the end-members that participated to the mixing process (Morgavi et al., 2013a; Perugini et al., 2013). This information can be obtained during the study of natural rock samples only if the mixing time was short. On the contrary, if the mixing time is too long the identification of the end-member compositions might be a non-trivial task since the most mafic composition tends to be erased in the mixing system in short times (Morgavi et al., 2013a; Perugini et al., 2013).

The Shannon entropy approach overcomes this limitation since it does not require information about the end-member compositions. The application of the Shannon entropy approach merely require knowledge of the theoretical hybrid composition to be eventually attained by the mixing system. If the mixing process developed for a sufficient long time, then this information can be acquired using the compositional histogram method (Perugini et al., 2004; 2013). Therefore, the Shannon entropy approach extends our capability in dealing with the complexity of magma mixing systems and their time evolution.

The above discussion is particularly relevant in petrological studies where the knowledge of the mobility of chemical elements is essential in order to evaluate quantitatively, through geochemical modeling, the processes involved in the genesis and evolution of a given set of rocks. The problem is that the mixing of silicate liquids does not require necessarily two end-members generated by completely different source rocks (e.g. mantle and crustal derived melts). This configuration is an extreme condition and probably only represents the extreme case. More generally, chemical gradients can be generated at any stage and at any time during the evolution of a magma body through any evolutionary process such as partial melting, fractional crystallization, or assimilation of host rocks (Perugini et al., 2006; De Campos et al., 2011; Perugini and Poli, 2012). This implies that melts with different compositions can coexist in the same system at any time. Further, given that the modulation of the compositional field is governed by chaotic dynamics, a fractal distribution of highly heterogeneous domains of melt is to be expected (Flinders and Clemens, 1996; Perugini et al., 2003; De Campos et al., 2011). As noted above, in principle, these melts have the potential to mix at any time. The result is the development of a diffusive fractionation of chemical elements due to their different mobility during melt advection by chaotic flow fields (Morgavi et al., 2013a; Perugini et al., 2012, 2006). This results in non-linear relationships between couples of chemical elements that violate the classic assumption that magma mixing should produce linear patterns of data points in inter-elemental binary plots (Perugini et al., 2006; De Campos et al., 2011; Morgavi et al., 2013b). Therefore, the use of the two end-member mixing equation is no longer valid, for example, to estimate the hybrid composition and the initial proportions of end-members. The combined use of Shannon entropy and Relaxation of Concentration Variance methods overcome these limitations because they are solely based on the fact that mixing processes are governed by chaotic dynamics. This allows us to address directly the non-linear complexity of magma mixing of silicate liquids providing a vital step towards a more complete view of this important petrogenetic process.

The combined application of the Shannon entropy and the Relaxation of Concentration Variance methods also can be used as a geochemical chronometer to measure the duration of mixing processes in volcanic rocks. As magma mixing is likely one of the main processes triggering explosive eruptions, their use might unravel unprecedented information about the timing of these natural events providing a further diagnostic tool to infer the timing of eruptions. Both methods show a clearly predictable time dependence; used jointly they can be employed to crosscheck the relative time estimates to provide more robust magma mixing timescales. Their application as volcanic chronometers, however, requires the construction of calibration curves relating Shannon entropy to the concentration variance and time. The required calibration can only be gathered by means of experimental time series of magma mixing. The recent development of highly accurate experimental devices for magma mixing (De Campos et al., 2011; Perugini et al., 2012) allows us today to obtain information about the development of magma mixing and its timescales that would have been unimaginable only ten years ago.

The above considerations are clearly valid within the boundary conditions of the numerical simulations and the high-temperature experiments used to test our conceptual method. In particular, the methods have been proven to be applied successfully when the mixing process occurred between degassed silicate liquids in Newtonian rheological conditions and with few or no crystals. Therefore, the method can be applied to volcanic rocks for which these conditions can be inferred. Although the Newtonian rheological behavior of magmas has been suggested to exist even at crystal fractions up to 20-25% (Caricchi et al., 2007; Mader et al., 2013), there are still several unknowns (i.e. the role of dissolved fluids and crystals) to be considered before our method can be applied to significantly crystal-bearing magmas at depth.

Further works focused on the development of high-temperature experiments of magma mixing using melts with different crystal contents and $CO_2/H_2O$ contents, as well as the study of natural outcrops showing similar properties will allow to calibrate the conceptual models presented here to a larger spectrum of natural systems and will potentially open a new window on igneous systems whose investigation by direct observations remains at present impossible.

**Acknowledgments**

This work was funded by PRIN2010-MIUR grant and the European Research Council Grant ERC-2013-CoG No. 612776 – CHRONOS (DP) and the ERC-2009-AdG No. 247076 (DBD). FAPESP provided a Visiting Professor grant regular project 2015/118500 and thematic project (2009/50493-8) (CDC). Ewa Slaby and an anonymous reviewer are gratefully thanked for their constructive comments.

**Figure captions**

Figure 1: Numerical simulations of fluid mixing at different mixing times: a) initial configuration (t=0) and after $t$=10 (b), $t$=20 (c) and $t$=30 (d) iterations of the numerical scheme. The panels show the evolution of the mixing system for an element with diffusion coefficient $D$=0.3. The line α–β represents the transect along which the compositional variability in the mixing systems is monitored in time.

Figure 2: a-d) Compositional variation ($C_x$) of the numerically simulated mixing system along the α–β transect (Fig. 1) for an element with diffusion coefficient $D$=0.3 at different times. On the top of each compositional transect a "bar-code" (BC) representation of the amount of the hybrid composition in the mixing system is reported at different times. The grey bands in the plots mark the location of the hybrid composition and are reported as a range to include ± 10% variability in the hybrid neighborhood (see text).

Figure 3: (a-c) compositional variability of the numerically simulated mixing system at the same number of iterations ($t=10$) for three elements having different $D$ values. Left panels: fluid mixing domain; the black line represents the transect along which the compositional variability in the mixing systems is monitored; right panels: compositional variation ($C_x$) along the transects highlighted on the left panels; on the top of each compositional transect a "bar-code" (BC) representation of the amount of the hybrid composition in the mixing system is reported. The grey bands in the plots mark the location of the hybrid composition and are reported as a range to include ± 10% variability in the hybrid neighborhood (see text).

Figure 4: Optical images of experimental samples at mixing time $t = 5$ min (a), $t = 20$ min (b) and $t = 120$ min (c). The white arrow in (a) indicates the direction of motion of the mafic melt during centrifuge rotation. Panels (d), (e) and (f) report the enlargement of sections marked in (a), (b), and (c) with white dashed lines, respectively. These correspond to the interface between the two melts where most of the mixing process occurred and where geochemical analyses have been performed (along dotted lines).

Figure 5: Compositional variation of representative major elements along the interface between the mafic and felsic melts (Fig. 4) for the 5 min (a), 20 min (b) and 120 min (c) mixing experiments. In the figure, concentrations of initial mafic, felsic and hybrid melts are also marked in different gray shades. These are reported as a range to include analytical uncertainties (Perugini et al., 2013).

Figure 6: a) Representative compositional series used to illustrate the generation of the point distribution of the hybrid data and the relative bar-code visualization. The grey area in the graph reports the interval used to identify the concentration of hybrid data points (i.e. ±10% of the value of the hybrid concentration; see text); b) distribution of the data points along the compositional series in (a) having attained the hybrid concentration (±10%; see text); c) bar-code representation of the hybrid distribution along the compositional series (a).

Figure 7: Example of Shannon entropy determination for the compositional transects reported in Fig. 2b. A grid is placed on the bar-code representation of the amount of hybrid compositions at different mixing times. This way the Shannon entropy is measured.

Figure 8: Representative variation of Shannon entropy as a function of mixing time ($t$) for the numerically simulated mixing system. Dashed lines represent the linear fitting of the rectilinear portions of the graphs for elements with different $D$ values.

Figure 9: Variation of Shannon entropy as a function of mixing time ($t$) for the experimental mixing system. Dashed lines represent the linear fitting of the rectilinear portions of the graphs for the different major elements. The 120 min experiment was not included in the linear fit.

Figure 10: Variation of concentration variance with mixing time for: a) three chemical elements simulated numerically having different values of $D$; b) three representative major elements analyzed on the experimental samples. Data for each element are fitted using [Eq. 2] and the Relaxation of Concentration Variance parameter *R* (*RCV*) is reported for each fitting.

Figure 11: Variation of the Relaxation of Concentration Variance (RCV) as a function of the rate of increases of entropy (*dS/dt*) for the numerically simulated mixing system. RCV values are taken from Morgavi et al. (2013a). Best-fit line and relative equation are also reported in the graph. Error bars represent uncertainties due to the extrapolation of *dS/dt* and RCV values by fitting the entropy vs. time (Fig. 8) and concentration variance vs. time relationships (Fig. 10a), respectively.

Figure 12: Variation of the Relaxation of Concentration Variance (RCV) as a function of the rate of increases of entropy (*dS/dt*) for the experimental mixing system. RCV values are taken from Perugini et al. (2013). Best-fit line and relative equation are also reported in the graph. Error bars represent uncertainties due to the extrapolation of *dS/dt* and RCV values by fitting the entropy vs. time (Fig. 9) and concentration variance vs. time relationships (Fig. 10b) (Perugini et al., 2013), respectively.

**Figure 1**

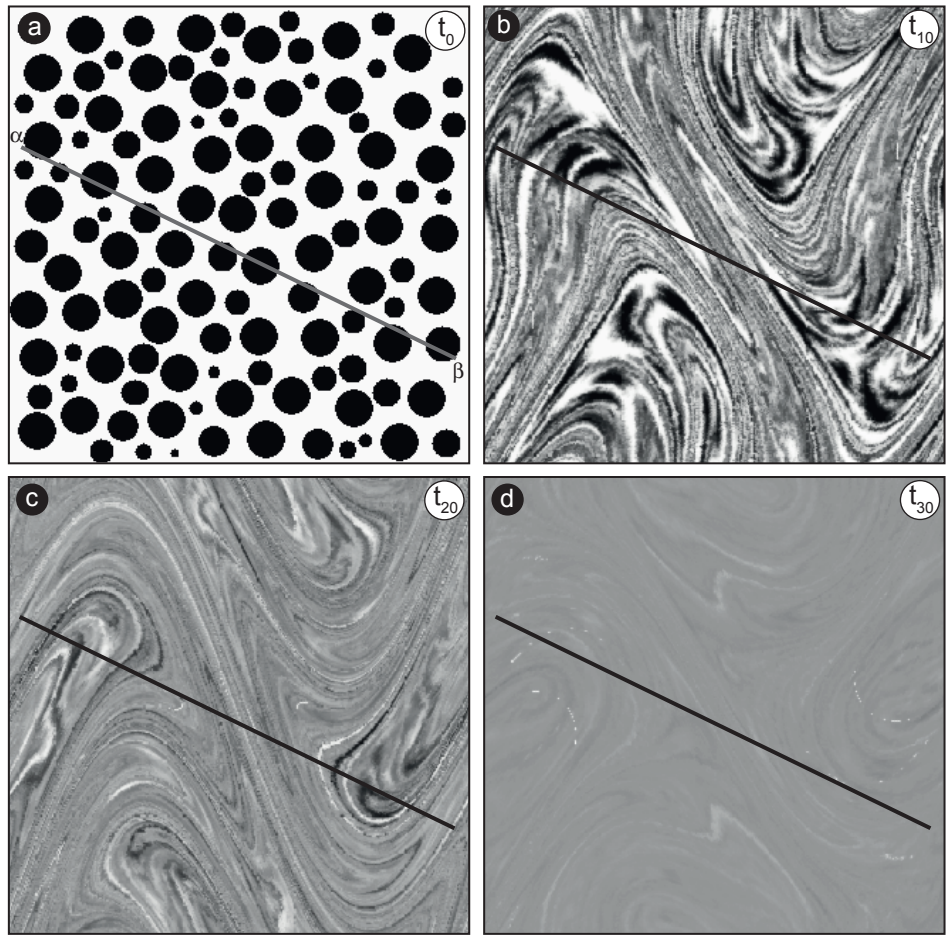

Figure 1

Figure 2

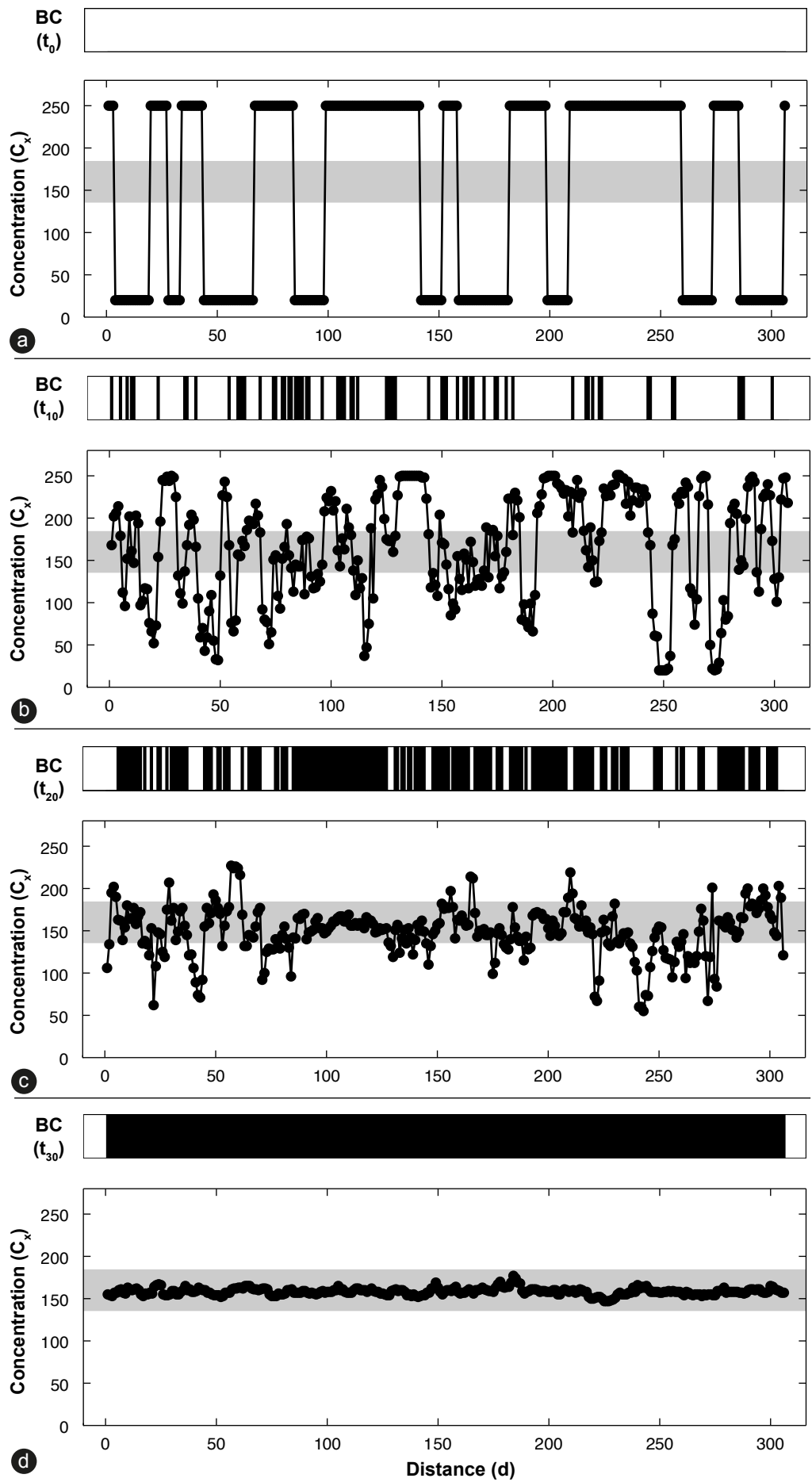

Figure 2

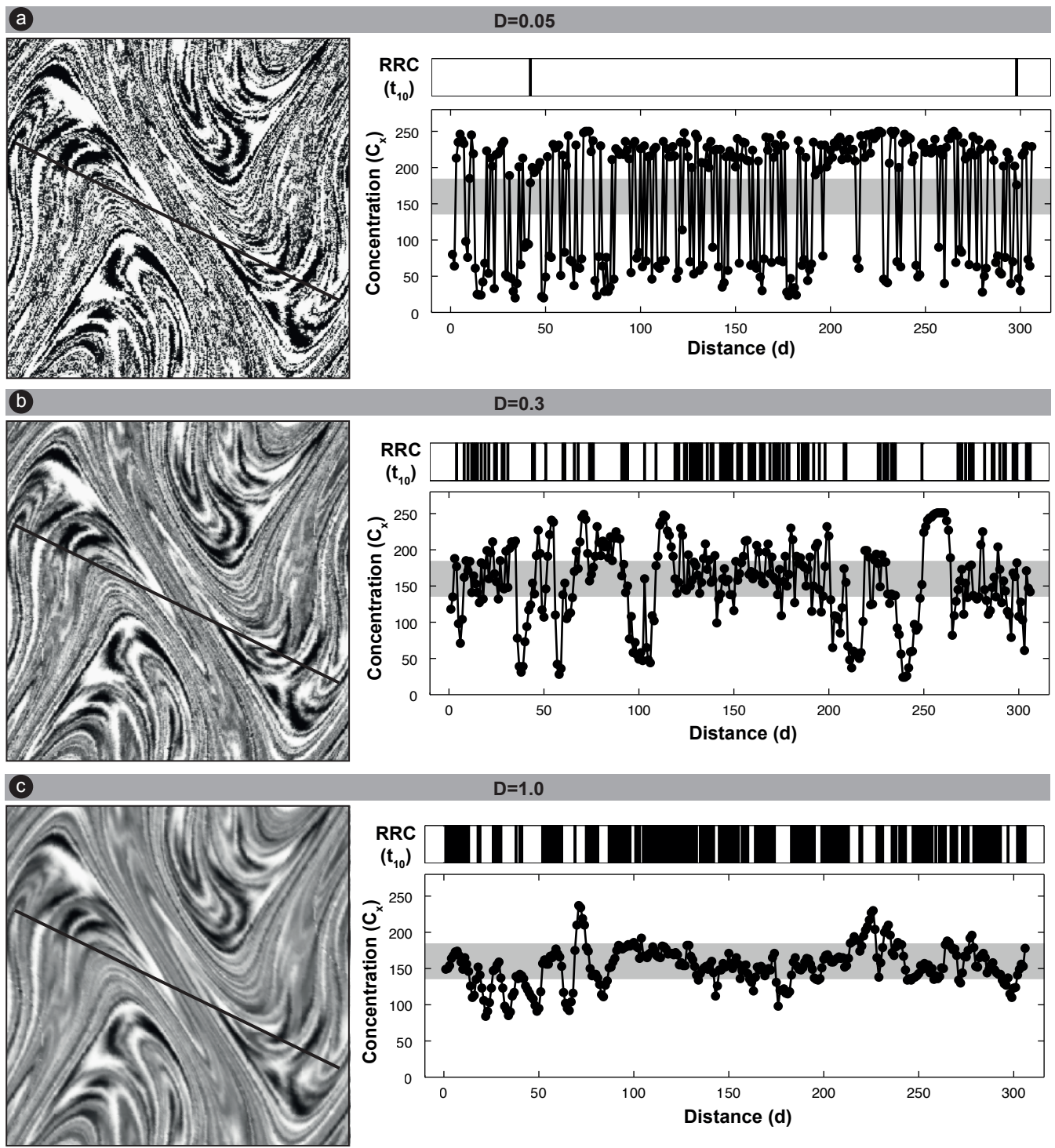

Figure 3

**Figure 4**

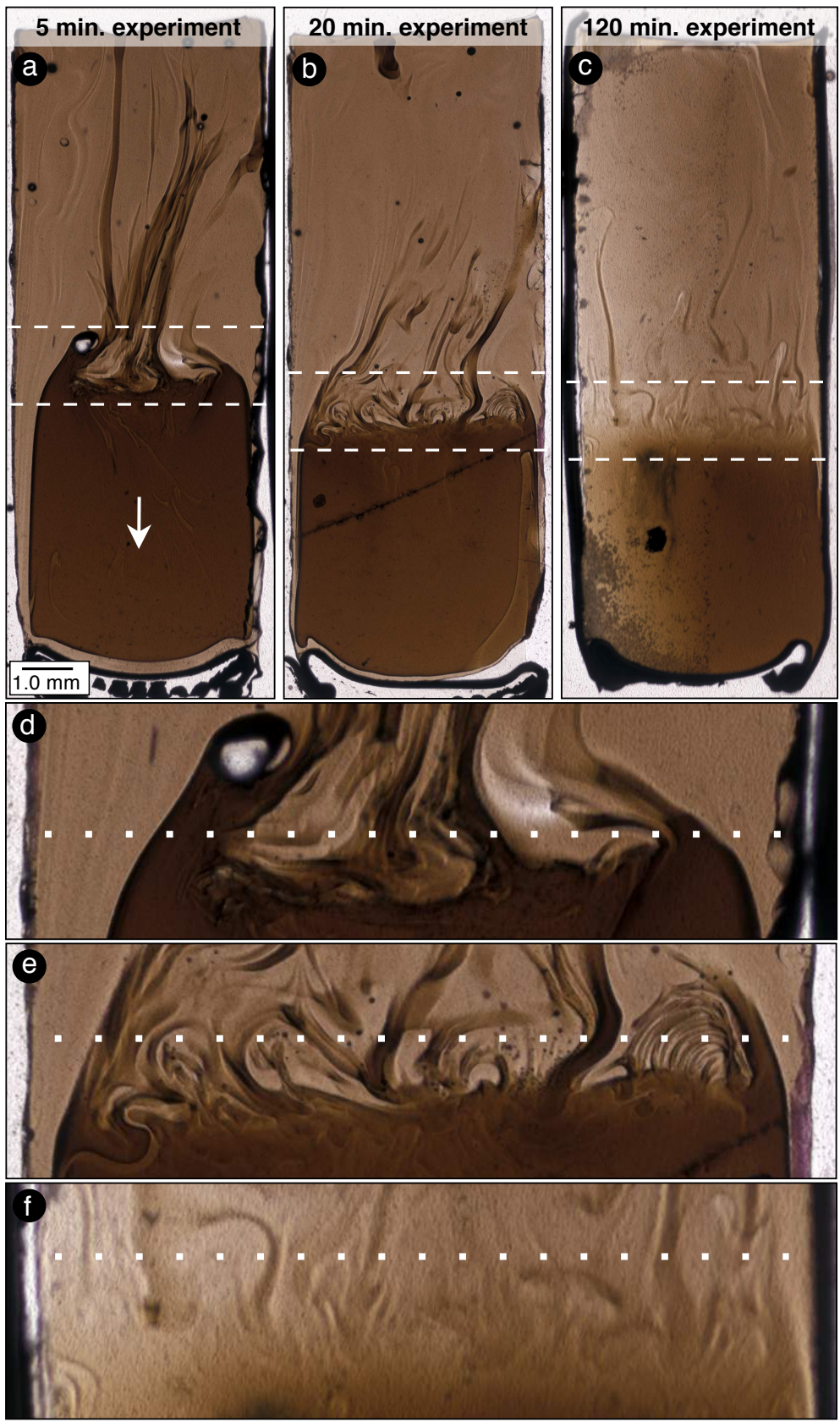

Figure 4

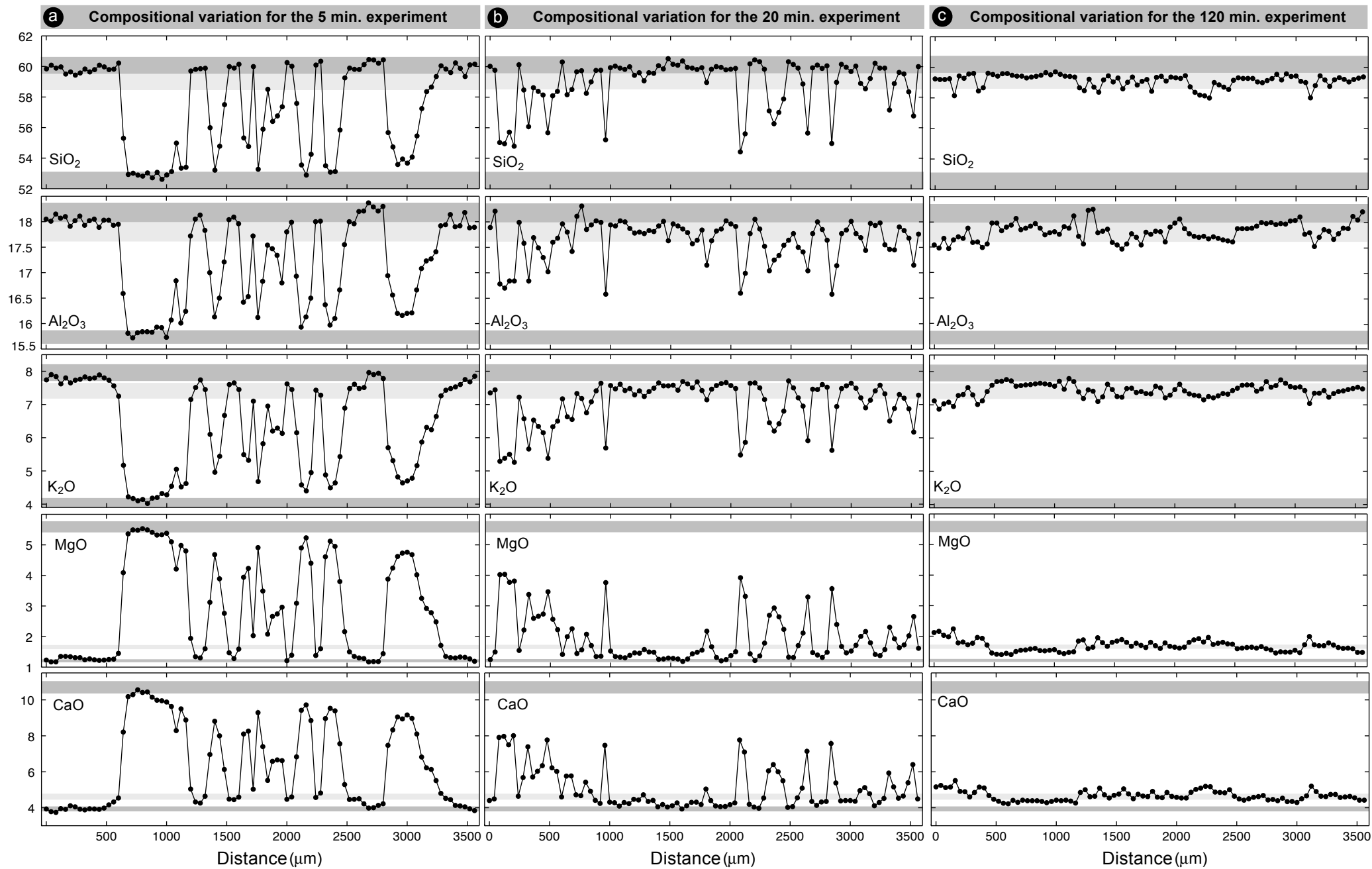

Figure 5

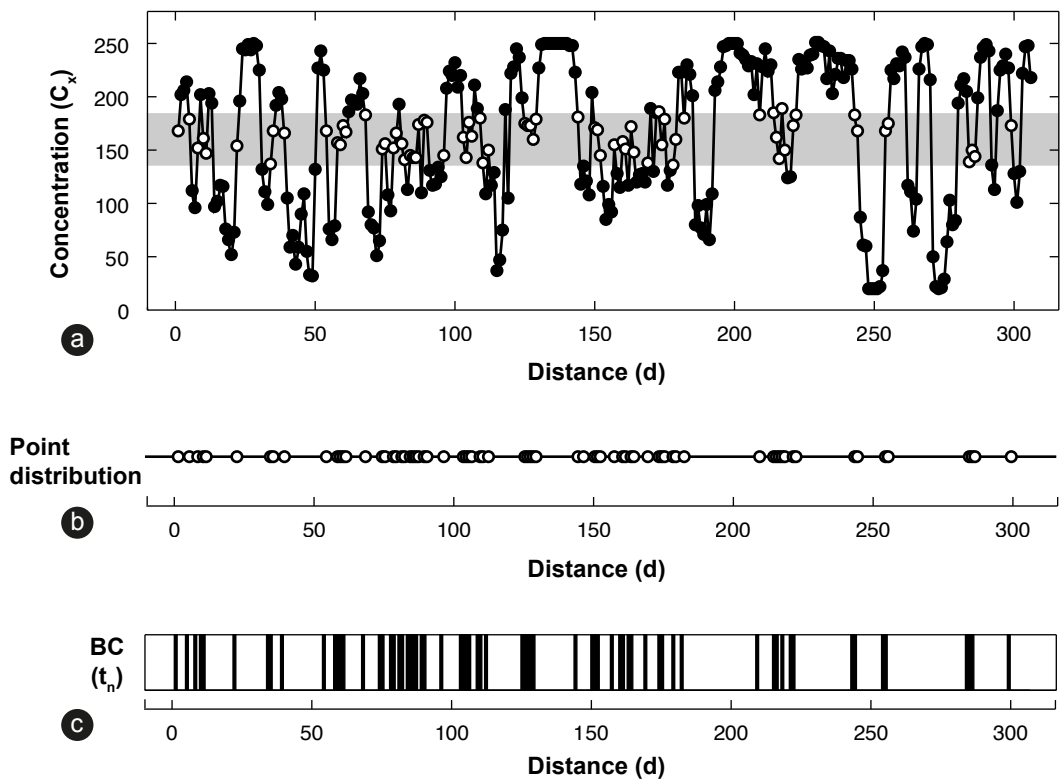

Figure 6

**Figure 7**

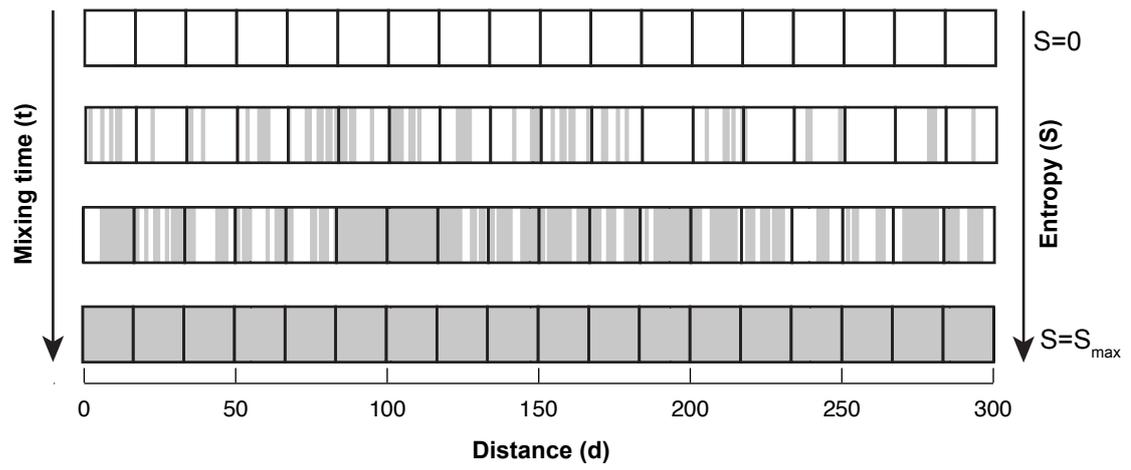

Figure 7

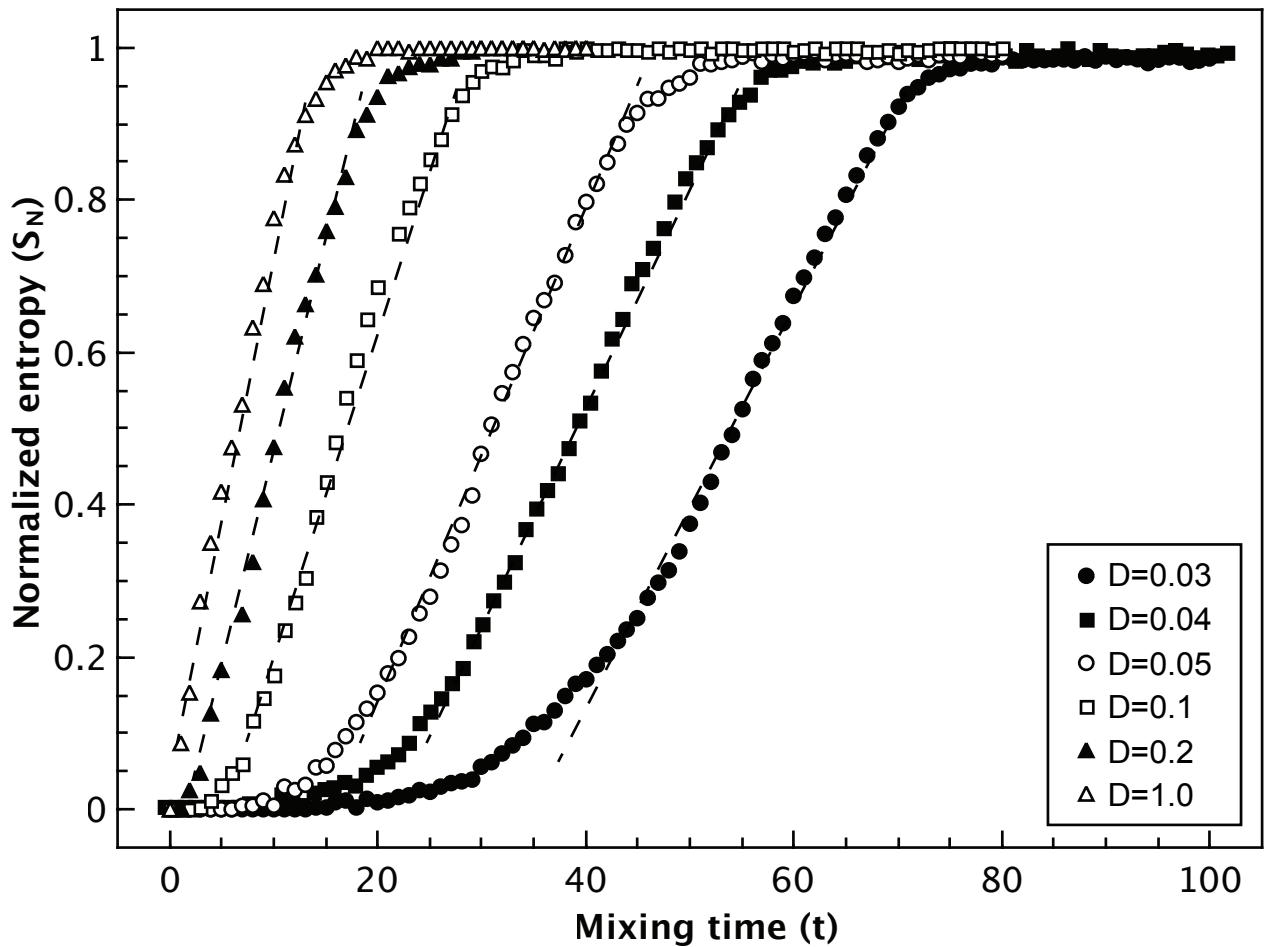

Figure 8

**Figure 9**

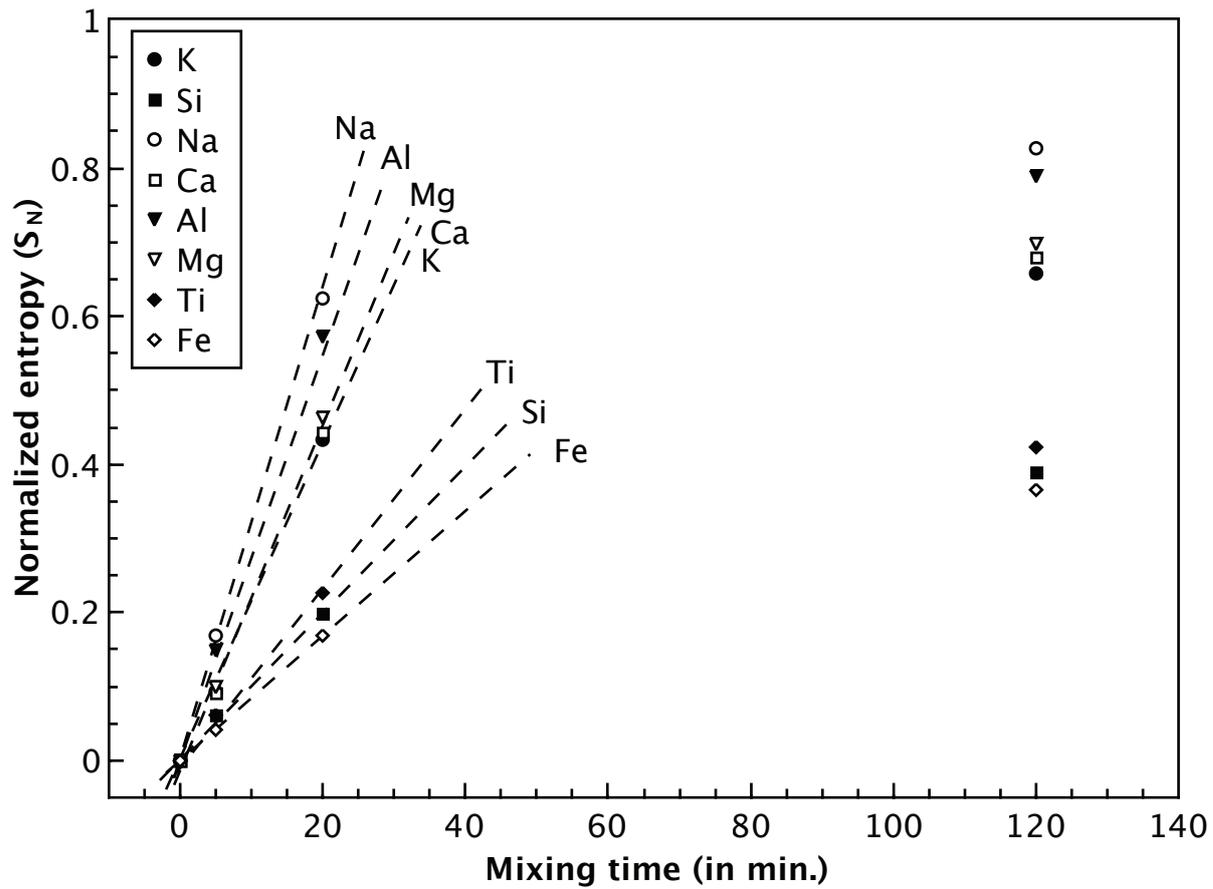

Figure 9

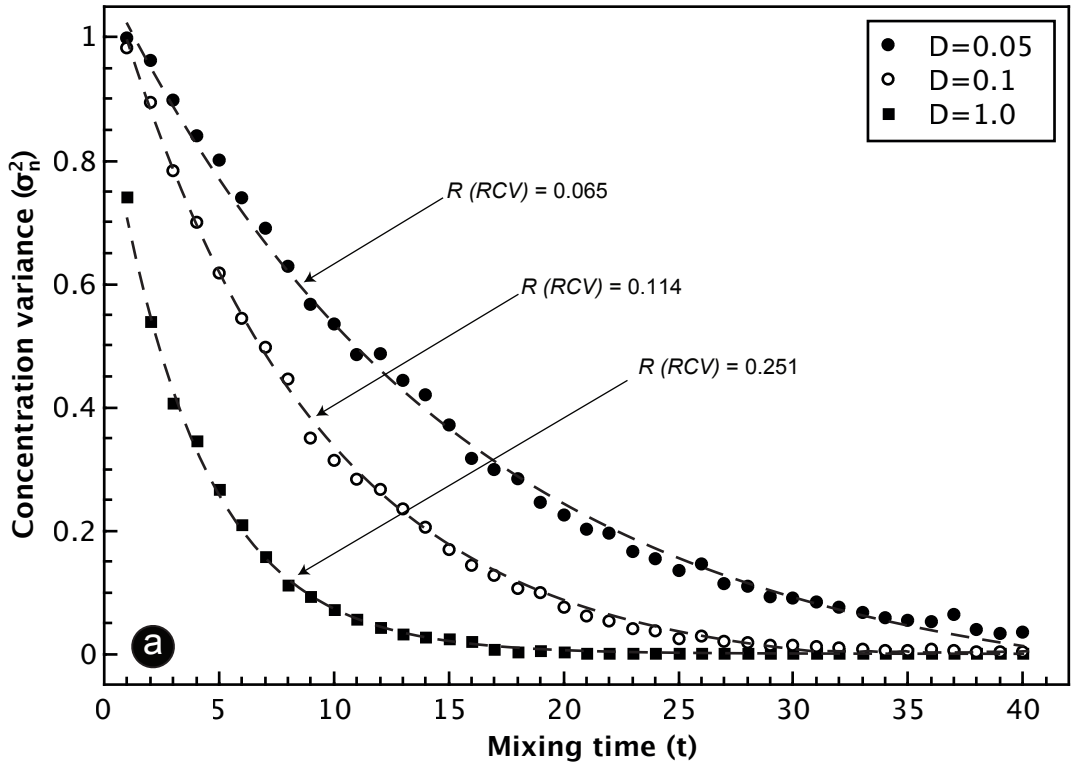

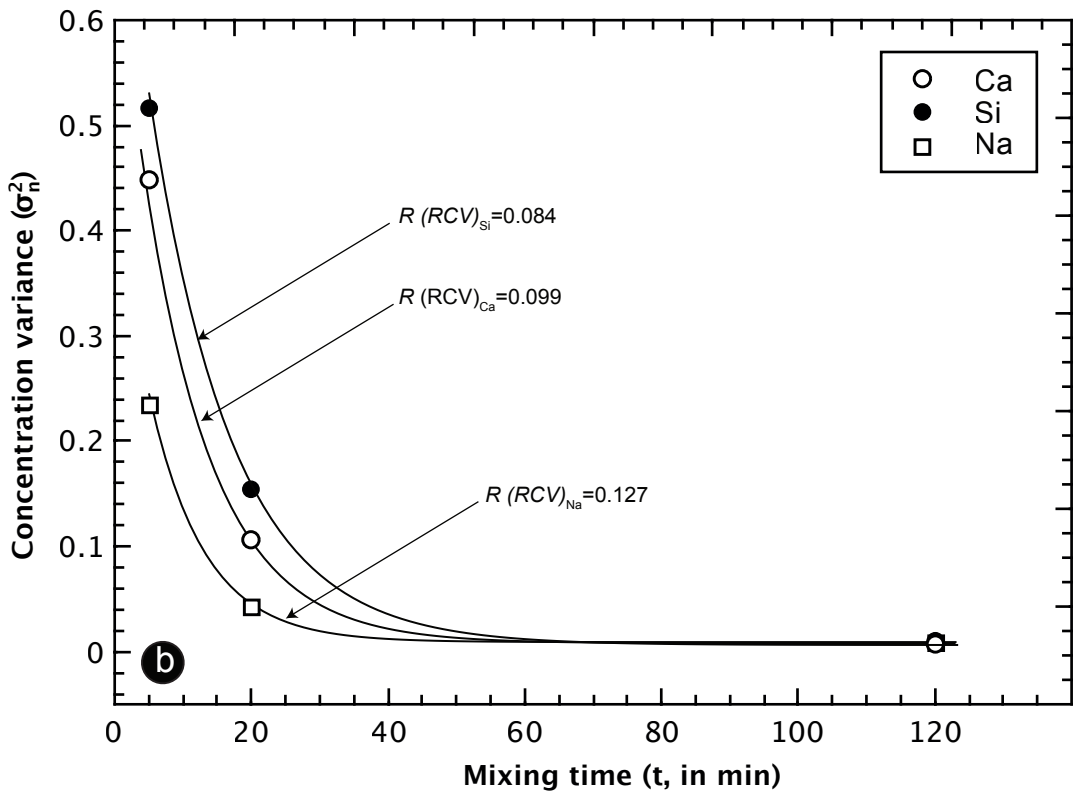

Figure 10

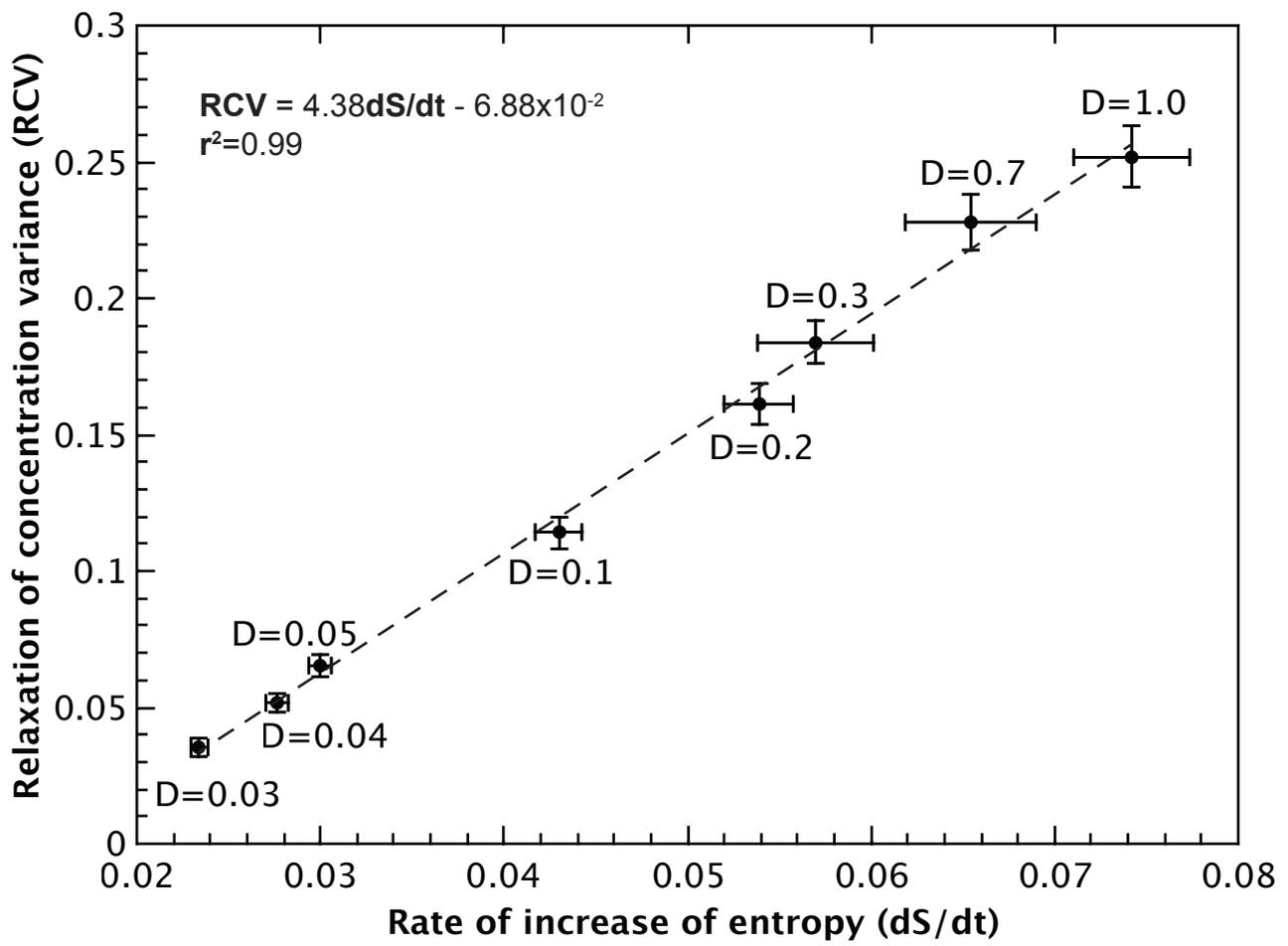

Figure 11

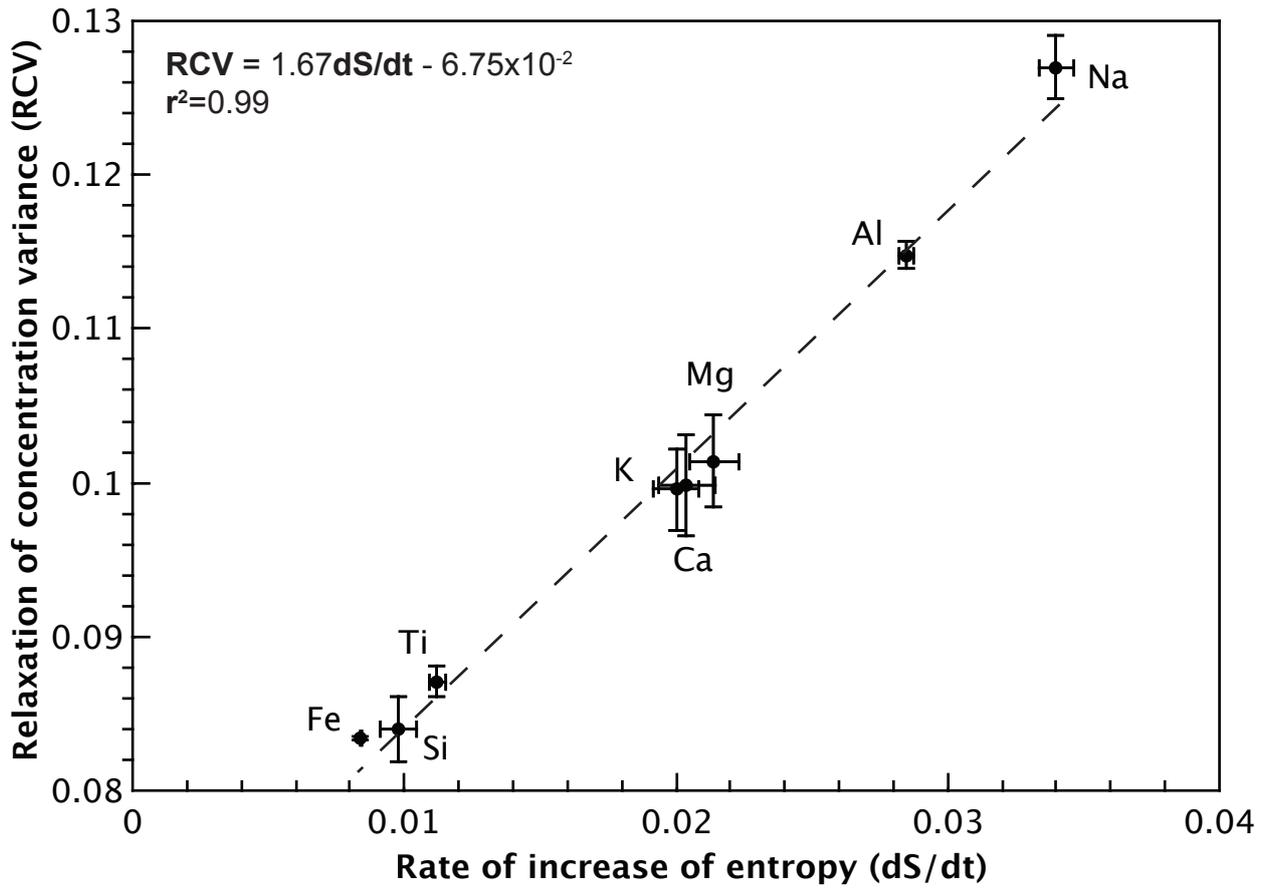

Figure 12